\documentclass[prb, twocolumn, letter, showpacs]{revtex4}

\usepackage{graphicx}
\usepackage{calc}
\usepackage{color}

\newlength{\halfwidth}
\newlength{\quarterwidth}
\newlength{\multiwidth}
\begin{document}

\title{Avalanches Injecting Flux into the Central Hole of a  Superconducting MgB$_2$ Ring}
\date{\today}

\author{{\AA}ge Andreas Falnes Olsen} \email{a.a.f.olsen@fys.uio.no}
\author{Tom Henning Johansen} \email{t.h.johansen@fys.uio.no}
\author{Daniel Shantsev} \affiliation{Department of Physics}
\affiliation{Center for Advanced Materials and Nanotechnology,
  University of Oslo, P.O. Box 1048 Blindern, 0316 Oslo}
\author{Eun-Mi Choi} \author{Hyun-Sook Lee} \author{Hyun Jung Kim}
\affiliation{National Creative Research Initiative Center for
  Superconductivity, Department of Physics, Pohang University of
  Science and Technology, Pohang 790-784, Republic of Korea}
\author{Sung-Ik Lee} \affiliation{National Creative Research
  Initiative Center for Superconductivity, Department of Physics,
  Pohang University of Science and Technology, Pohang 790-784,
  Republic of Korea} 

\begin{abstract}

Magneto-optical imaging was used  to observe dendritic flux avalanches  connecting the outer and inner edges of  a ring-shaped superconducting MgB$_2$ film. Such avalanches create  heated channels across the entire width of the ring, and inject large amounts of flux into the central hole. By measuring the injected flux and the corresponding reduction of current, which is typically 15 \%, we  estimate the maximum temperature in the channel to be  100~K, and the duration of the process to be on the order of a microsecond. Flux creep simulations  reproduce all the observed features in the current density before and after injection events.

\end{abstract}



\maketitle

\section{Introduction}

In recent years it has become clear that instabilities are commonplace when magnetic flux penetrates superconductors. In particular, magnetic flux avalanches can abruptly penetrate thin superconducting samples in the form of  tree or fingerlike patterns known as dendrites. Such avalanches were first observed in YBa$_2$Cu$_3$O$_x$, where they must be triggered by a sudden point-like heat-pulse \cite{Leiderer1993,Bujok1993}. Since then spontaneous dendritic avalanches have been  observed \cite{Duran1995,Rudnev2003,Welling2004,Wimbush2004,Rudnev2005} in  Nb, Nb$_3$Sn, YNi$_2$B$_2$C and NbN, as well as in patterned Pb thin films \cite{Menghini2005}. The recent interest in the phenomenon was largely triggered by the discovery \cite{Johansen2001} that in MgB$_2$ films  the avalanches are ubiquitous  below a threshold temperature $T_{th} \sim 10$~K. The dendrites disrupt electrical current flow and limit the overall current capacity  of superconductors, \cite{PhysRevB.65.064512} and are thus harmful for prospective applications. On the other hand they also represent fertile ground for fundamental research as there are  many interesting challenges in understanding their nucleation and growth.

The dendritic instability is  now believed to be of  thermo-magnetic origin. Motion of flux in superconductors releases heat and increases the temperature. Flux motion is facilitated by the increase in temperature,  so more flux moves which then leads to even more heating. If the released heat is not transported away fast enough an instability develops whereby large amount of magnetic flux rushes into the sample. This mechanism was originally invoked to explain global  flux jumps where the entire sample is filled with a sudden burst of magnetic flux and all magnetisation is lost  \cite{Swartz1968,Mints1981,Mints1996}.  However, the mechanism has recently also been shown to give localised, fingering instabilities.  \cite{Aranson2005,Denisov2006} Indeed, the instability threshold field predicted from this model  was recently \cite{denisov:077002} shown to agree quantitatively  with measurements in both Nb and MgB$_2$ films. 

The dynamics of dendritic avalanches is extremely challenging to probe experimentally because of the short times involved. Typically the entire process of nucleation and growth of a single flux dendrite lasts a few hundred nano-seconds as measured using a sophisticated magneto-optical (MO) set-up \cite{Leiderer1993,Bujok1993}. It is particularly difficult to measure the expected increase in temperature associated with the avalanches.  On the other hand, it is possible to estimate the temperature in the dendrite indirectly by designing a suitable geometry where dendritic avalanches can leave a long-lasting, temperature dependent imprint.  In particular, in superconducting rings dendrites may span the entire ring width and thus create a heated channel that connects the inner and outer edges. During the brief time that this channel is open it injects flux into the otherwise screened central hole in a process which may be described as magnetic perforation.  This process depends on the temperature in the channel, which may then be inferred from measurements of the flux contained in the hole. 

In this paper we report a detailed study on the perforation process in a superconducting MgB$_2$ ring using MO imaging. The technique, described in section \ref{sec:experimental},  provides us with a detailed map of the flux distribution over a sample-wide area. In section \ref{sec:results}  we present our experimental results, showing  the evolution of the  flux $\Phi_{r < r_p}$ contained within the screened region $r < r_p$ during field ramps at different temperature, as well as the current distribution in the ring.  $\Phi_{r < r_p}$ displays  abrupt jumps when dendrites cross the ring. From a simple model presented in section \ref{sec:peaktemperature} which connects the temperature evolution in the heated channel and  the evolution of the  current during perforations  we  estimate the peak temperature during flux injection. We find that this temperature is  $2.5 T_c$, and that the  duration of the flux injection process is on the order of one $\mu$s. Finally in section \ref{sec:simulations} we propose a numerical model  to describe how the perforation affects the current distribution in the ring. 

\section{Experimental set-up}
\label{sec:experimental}

A MgB$_2$ ring was  prepared by pulsed laser deposition (PLD) on a sapphire substrate.
Details of the manufacturing process and typical properties of the
superconductor can be found elsewhere \cite{Kang2001,Choi2005}. Using photolithography the
superconducting film of thickness $d=0.5$~$\mu$m was  patterned into a
ring with inner radius $r_{0} = 1.10$~mm
and outer radius $r_{1} = 2$~mm.

The magneto-optical indicator was a  ferrite garnet film (FGF) of thickness 5~$\mu$m grown
on a gadolinium gallium garnet (GGG) substrate and covered with an Al mirror. Monodisperse beads (diameter 3.5~$\mu$m) were  sprinkled on the sample  to ensure a small and uniform distance between the superconductor and  the Al mirror prior to positioning of the indicator. This precaution is necessary since metallic layers are known to suppress dendritic avalanches \cite{Baziljevich2002,Choi2005} in MgB$_2$.  The
entire stack was mounted in an optical cryostat (a liquid helium flow cryostat from Oxford Instruments) and
cooled to low temperature for observation in a polarisation
microscope. We ensured sensitivity to field direction by setting the polariser - analyser angle $\alpha$  a few degrees away from full extinction. Images were captured with a QImaging Retiga Exi CCD camera
with 12-bit pixel depth, controlled from a computer. The coil current
was controlled from the same computer by means of  a DAC card used to generate a control voltage to the current source. The computer also communicated with an external voltmeter (Hewlett Packard 34401A), which was used to monitor the coil current. The experiments were then conducted by slowly ramping the applied field from 0 to
25~mT while recording $50 - 200$ images at regular intervals. Each time an image was recorded, the coil current was read and written to a log file before proceeding with the field ramp.

We calibrated image brightness to field by capturing $20-30$  images at low temperature and  a point away from the sample, while ramping the
applied field from 0 to $40$~mT. The recorded applied field and the image brightness were used to fit  a second-degree polynomial $I(B) = a_2 B^2 + a_1 B + a_0$ relating magnetic field to the image intensity. This polynomial form follows from Malus law for light intensity in a polarisation microscope,

\begin{equation}
  I = I_{0}\sin^{2} (\theta + \alpha) + I_{b}
\end{equation}

along with the phenomenological linear relation $\theta \propto B$ for the Faraday rotation at small angles. 

The calibration was performed pixel-by-pixel, which is necessary  to account for uneven illumination in the field of view. The raw images were then processed in two stages: firstly, each image was
converted to a field map using the obtained values of the fitting coefficients $\{ a_i \}$, and secondly the flux $\Phi_{r<r_p}$  was computed by summing the field at each pixel contained within the relevant region. 

 Sensor noise leads to observable noise in the field values computed for each pixel, but summing over many thousand pixels to obtain the flux in a region effectively reduces random noise to insignificant levels. A more serious concern in quantitative magneto-optical measurements is systematic errors (see ref \onlinecite{Jooss2002}), particularly the existence of in-plane domains in indicator films which show up as triangular features of brighter and darker areas. The domain boundaries move easily in response to changes in temperature and applied field, and therefore result in a changing background.  We minimised their influence  by mounting the indicator very carefully on the sample to avoid mechanical stresses, and by applying a weak in-plane field. Despite these precautions domain splitting did occur in the experiments presented here. This was observed as dark regions slowly moving into the field of view during the field ramps, thus leading us to calculate a too low field value in the affected pixels. Fortunately, visual inspection of the images reveals that only a small part of the interesting region is affected, never more than $\approx 10$~\% of the total area. 

\section{Experimental Results}
\label{sec:results}

\begin{figure*}
  \includegraphics[width=\halfwidth]{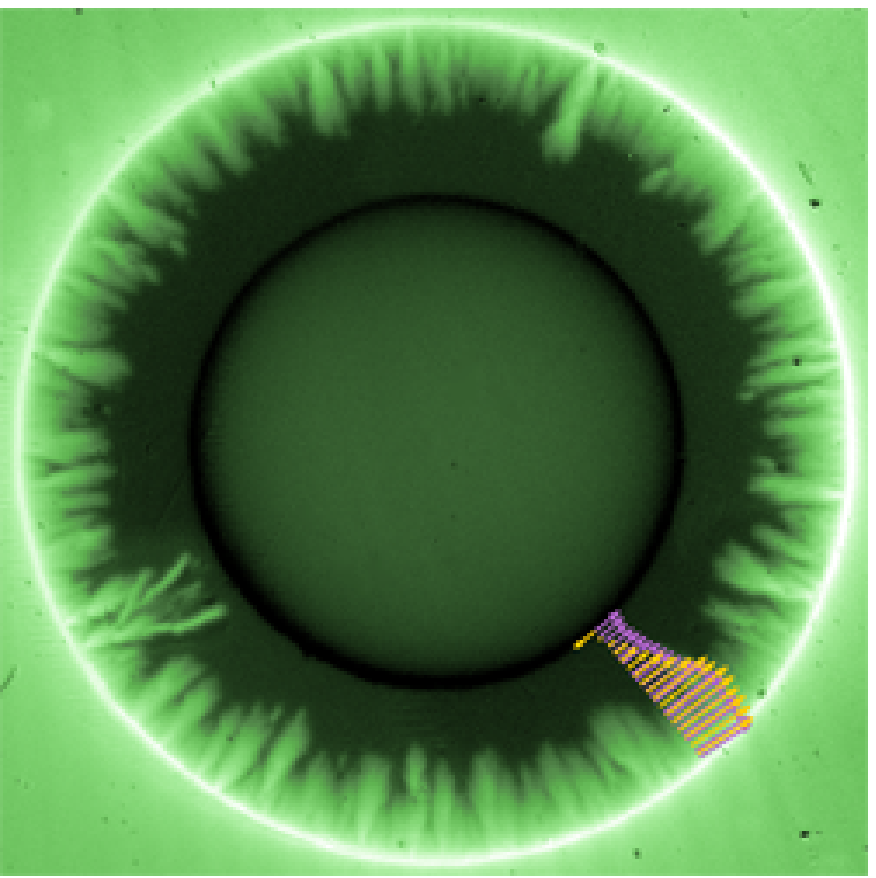} 
  \includegraphics[width=\halfwidth]{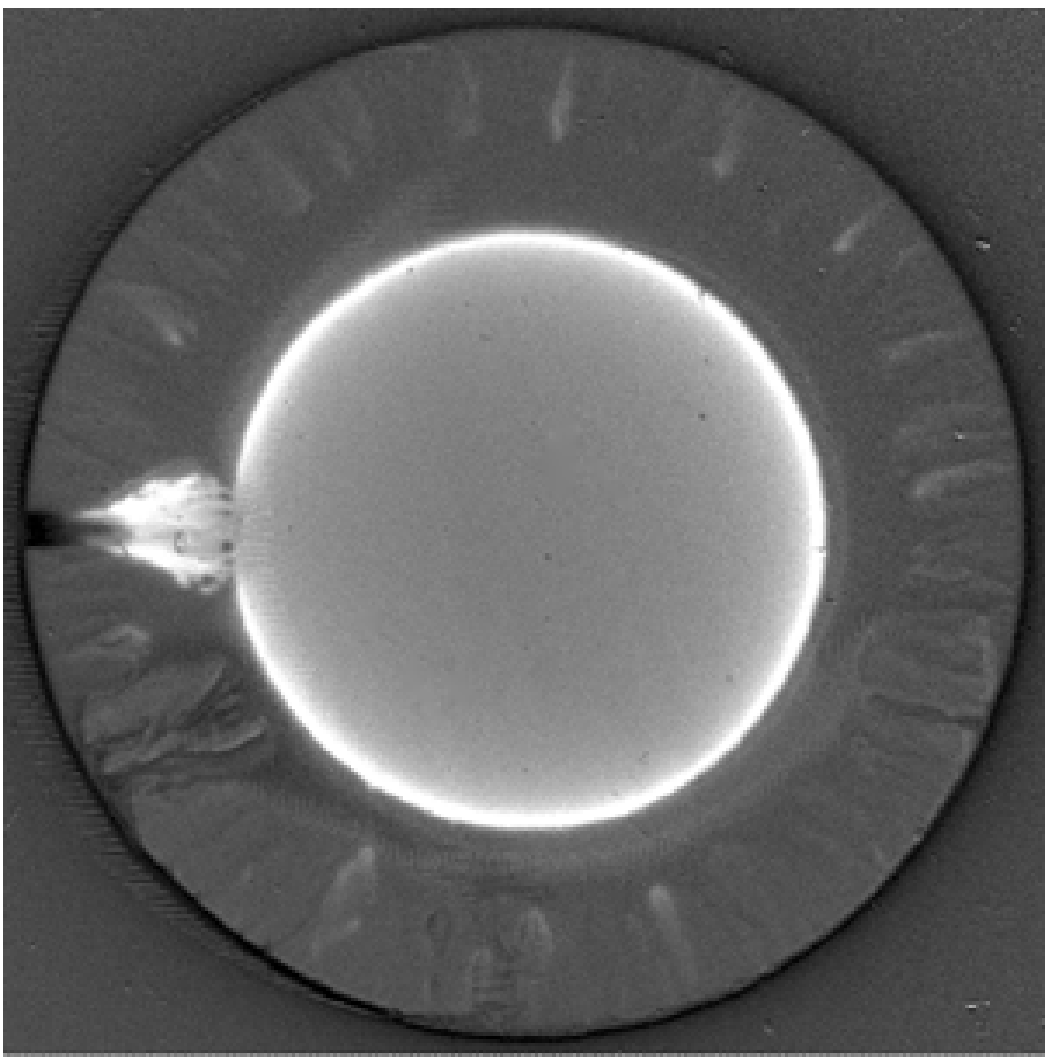}
  \caption{(left) A field map of the sample at $T=6.2$~K and $B_a = 13$~mT.   (right) The difference in field between the map shown here and the subsequent map in the sequence after a dendrite has perforated the ring and injected a large amount of flux into the central hole. A corresponding reduction in the field is seen around the outer edge. In the superconducting material there is very little change in the field. The arrows in the field map show the current before and after the perforation calculated from the field maps.}
  \label{perforationimages}
\end{figure*}

Figure \ref{perforationimages} (a) shows an example field map recorded at $T = 6.2$~K after applying a perpendicular field to the zero-field-cooled sample. Notice the negative field at the inner ring edge which is the accumulated stray field from the supercurrents. In the lower left quadrant of the ring we observe one of the   relatively few, but also relatively large,  dendrites formed at this  temperature. In fact, the image shown here was the last taken prior to the nucleation and growth of a second dendrite which spanned the entire ring width and injected large amounts of flux into the central hole. This perforation event is illustrated in figure \ref{perforationimages} (b) showing the difference between the frame in (a) and the subsequent frame taken at a slightly larger field. The increase in flux inside the central hole, and the accompanying reduction near the outer edge, are evident: the dendrite brings large amounts of flux from  outside the sample into the central hole. Remarkably, as dramatic as the perforation event is, the flux distribution inside the superconducting material is left largely unchanged in the process. 

Well below the threshold temperature $T_{th}$ we observe more frequent dendritic avalanches. At $T = 3.8$~K the first dendrite appears at roughly $B_a=5$~mT. As reported previously \cite{article9} the size of individual dendrites increases with the applied field, until they span the entire ring and cause perforations.


For low applied fields, before the first perforation event occurs, the superconductor screens the area contained within $r_p$, a characteristic radius where the flux fronts from the outer and inner edges meet in a critical state superconducting ring. The flux inside this region, $\Phi_{r<r_p}$, is the sum of the contribution from the supercurrents $\Phi_{self}$ and the applied field $\Phi_{B_a} = B_a A_{r<r_p}$: 

\begin{equation}
  \label{eq:fluxformula}
  \Phi_{r<r_p} = B_a A_{r<r_p} + \Phi_{self}
\end{equation}

\begin{figure}
  \includegraphics[width=\halfwidth]{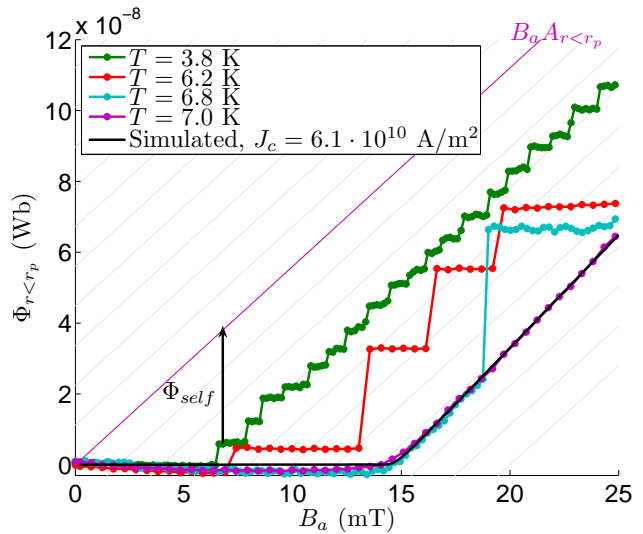}
  \caption{Measured flux within $r_p$  as a function of applied field showing crossover from complete screening to a linear increase following the applied field. The flux jumps are small and frequent at low temperature, becoming rarer and larger when the temperature approaches $T_{th}$. The oblique grid lines have slope $A_{r < r_p}$, with the line through the origin  representing the contribution to $\Phi_{r < r_p}$ from the applied field. The contribution to the flux from the ring currents, $\Phi_{self}$, is illustrated for one of the curves.  The solid black line is the result of a simulation with a Kim model for the critical current (see section \ref{sec:simulations}).  }
  \label{screenedflux}
\end{figure}

Provided no perforations have occurred and the applied field is below the field of full penetration $B_p$, this sum is exactly zero. Figure \ref{screenedflux}  shows the evolution of $\Phi_{r<r_p}$ during  ramps of the applied field at four different temperatures.  For the highest temperature $T=7$~K (above the threshold $T_{th}$ for the onset of dendritic avalanches in the present experiment) the flux remains zero until the applied field reaches $B_p$, after which it increases linearly with the applied field. The crossover results from the fact that the current in the superconductor has saturated and hence the term $\Phi_{self}$ remains essentially constant when the applied field is increased further. The light gray oblique lines in figure \ref{screenedflux} show slopes expected from the $B_a A_{r<r_p}$ term.  Guided by these lines one can see that the slope of the measured curve is slightly larger than expected if the supercurrents remained constant.  This means there is some relaxation of the supercurrent in the ring as the applied field increases. 

The central observation in this paper is seen in the curves obtained at  temperatures below $T_{th}$.   $\Phi_{r < r_p}$ is still zero up to a certain level, the perforation field, where it starts to increase in a step-like fashion, with sudden jumps followed by periods of screening. At $T = 3.8$~K there are  small and frequent jumps above the  perforation field, which in this  case is $\approx 6.5$~mT. The average slope of the flux curve  corresponds to the flux increase due to the applied field, but  $\Phi_{self}$ is much smaller than for the normal critical state above $T_{th}$. This means that the avalanches reduce the maximum total current which can flow in the ring. 

The curves at $T = 6.2$~K and $6.8$~K show similar behaviour, but with larger and fewer jumps and a slightly larger perforation field. In the latter case  the jump occurs well \emph{after} the field of full penetration has been reached.  

Clearly the flux jumps are accompanied by an abrupt reduction of the total ring current, with a  recovery taking place during the screening phases. However, it is not only the total current which is affected, but also the current distribution.  Using an efficient inversion scheme based on the Biot-Savart law which takes advantage of the fast Fourier transform \cite{roth:361}, we obtain  2-D current distributions  from our measured magnetic field maps. Superimposed on the field map in  figure \ref{perforationimages} (a) are two sets of  arrows showing the direction and density of current just prior to and just after a perforation. In the former case the supercurrent is qualitatively what we expect in a critical state ring, with a saturated current in the flux penetrated part of the sample and a smaller screening current in the Meissner state region. The perforation profoundly changes this distribution.  Near the outer edge there is a significant reduction in the current. Further from this edge, but within the flux penetrated region there is  no discernible  change in the current, while it is reduced throughout  the Meissner region. However, the greatest change occurs close to the inner edge, where  the current actually changes direction.

\begin{figure}
  \includegraphics[width=\halfwidth]{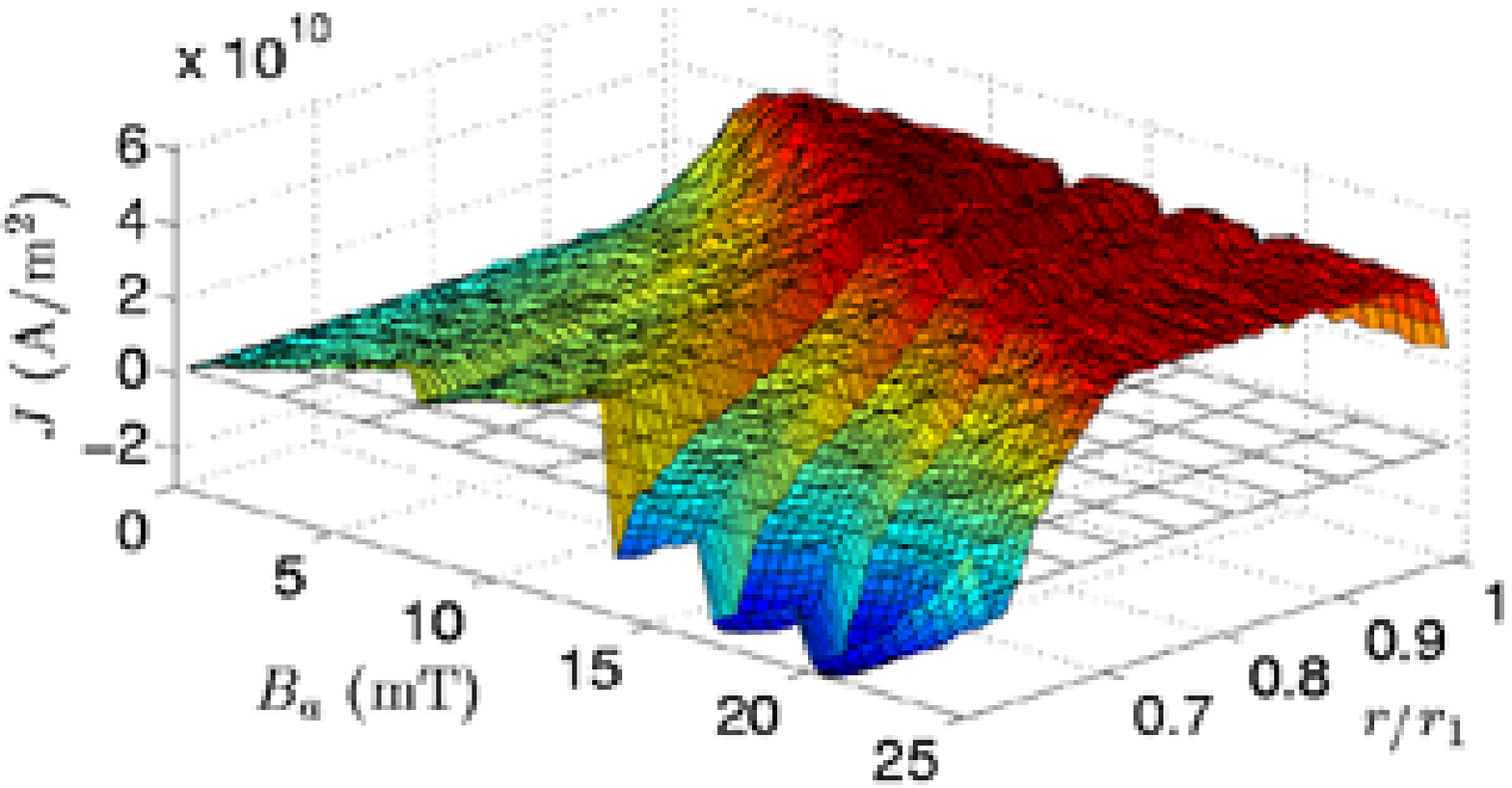} \\
  \includegraphics[width=\halfwidth]{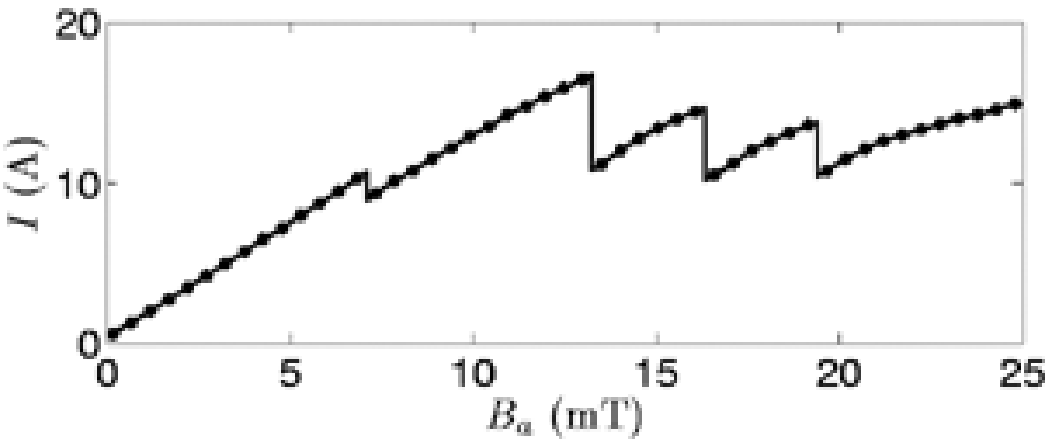}
  \caption{(top) The evolution of the current along the radial profile and for the same experiment shown in figure \ref{perforationimages}. (bottom) The total ring current obtained by integrating the current density across the ring width}
  \label{fig:currentevolution}
\end{figure}

All these features are typical after perforation events. For the particular image sequence under discussion one can see the complete evolution of the current in  figure \ref{fig:currentevolution}. Each of the four perforations that occurred in the sequence are visible as distinct reductions in the current level, with most of the change occurring near the inner edge. 

\section{Peak temperature and duration of the perforations}
\label{sec:peaktemperature}

The experiments give accurate measurements of the state just before and just after perforations. Despite the fact that the dynamics is far too quick for our experimental set-up, the fact that we \emph{can} observe the change in $\Phi_{r<r_p}$ allows us to make some qualified inferences about the details of the process. The starting point is to split the process into two  phases: (i) the \emph{nucleation and growth phase} when the dendrite tip rapidly moves from the outer to the inner edge creating a heated channel, and (ii) the \emph{flow phase} when the resistive channel  simultaneously impedes the supercurrents and injects flux into the central hole. We emphasise that during the first phase there should be little or no change in $\Phi_{r<r_p}$. Its duration is only determined  by the dendrite tip velocity. Ultrafast MO measurements \endnote{P. Leiderer, private communication} has been used to measure the tip velocity of dendrites in  MgB$_2$ yielding a value $v_{tip} = 10^5$~m/s. This tip velocity gives a crossing time $\delta t_{1} = 10^{-8}$~s in our sample. 

The second phase is more involved. For a narrow ring or a uniform current distribution the evolution of the current can be expressed as a balance between the voltage drop $RI$ across the heated channel  and the induced emf voltage $L \dot{I}$ due to the change in the ring current: 

\begin{equation}
  L \frac{dI}{dt} + R(t) I = 0
  \label{LR}
\end{equation}

$R$ is an ohmic resistance resulting either from a normal state resistivity above the critial temperature $T_c$, or a flux flow resistivity originating from moving vortices, and is non-zero  as long as the channel remains open. The inductance $L$ is a property of the ring, and is only determined by the geometry. The joule heating increases the temperature in the channel, a process which can be described by the heat equation

\begin{equation}
  \label{eq:heatequation}
  c \frac{dT}{dt} =  \rho J^2 - \frac{h}{d} (T - T_0)
\end{equation}

where we have ignored a term describing heat diffusion into the surrounding superconducting material. $c$ is the heat capacity, $h$ is the heat transfer coefficient to the substrate, and $T_0$ is the temperature of the substrate assumed to be equal to  the cold finger temperature. The two equations  are in fact coupled, since the current in the $\rho J^2$ - term evolves according to equation (\ref{LR}). Moreover, the resistivity is strongly temperature dependent below $T_c$, and hence the evolution of the current is influenced by the channel temperature. Physically, one can think of the process in the following way: as phase 2 commences the $\rho J^2$ term dominates over the heat removal term leading to an increase in the temperature. But this has two consequences, both of which tend to oppose a further increase in temperature: firstly, the heat removal term becomes larger for higher temperature, and secondly the current is reduced in the ring and hence the heat generation is reduced. At some point heat removal equalises the heat generation and a maximum is reached, after which the temperature decreases  until the current density drops below $J_c$. At this point the resistivity vanishes, and the flux injection halts.

 The fact that $c$, $R$ and $h$ are all temperature dependent means an analytic solution is difficult to find.  However, equation (\ref{eq:heatequation}) can be used to estimate the maximum temperature during phase 2 directly by putting $dT/dt = 0$.  Following ref \onlinecite{denisov:077002} we write the heat transfer coefficient  as $h(T) = h_0(T/T_c)^3$ and obtain the following equation for the maximum temperature $T_{max}$

\begin{equation}
  \label{eq:stationaryT}
  \frac{\rho J^2 d T_c ^ 3}{h_0} = T_{max} ^3 (T_{max} - T_0)
\end{equation}

 From   figure  \ref{fig:currentevolution} we find that at perforation the average current density over the whole ring  is  $J \approx 0.45 J_c = 2.7 \cdot 10^{10}$ A/m$^2$. The heat transfer coefficient   was estimated in reference \onlinecite{denisov:077002} to be $h_0 = 20$~kW/Km$^2$ on samples manufactured in the same way on the same sort of substrate as the current sample. If the temperature is moderately above $T_c$, the resistivity is independent of temperature in this material \cite{Eom2001,Kang2001,Kim2001} and can be taken as $\rho_n = 10^{-7}$~$\Omega$m. Finally, with $T_c = 40$~K we obtain a peak temperature during the flux injection process of $T_{max} = 105$~K.

We can now estimate the duration of phase 2 from equation (\ref{LR}). With the above temperature estimate in mind  $R$ is determined by the normal state resistivity,   $R = \rho_n w / A$. It follows that equation (\ref{LR}) describes a simple exponential decay of current, $ I(t) = I_{before} \exp (- \rho_n w / AL)$ with $I_{before}$ being the ring current just prior to the perforation. The area  $A=45 \cdot 10^{-11}$~m$^2$ is the radial cross section of our sample,  and $w$ is the width of the resistive region (the dendrite core). For the jump in flux $\Delta \Phi_{self} = (I_{after} - I_{before})L$, we obtain the timescale for the flux flow phase as

\begin{equation}
  \label{eq:phase2time}
  \delta t_2 =  \frac{L A}{w \rho} \ln \left( \frac{1}{1 - \Delta \Phi_{self} / \Phi_{self}} \right)
\end{equation}

The plots in figure \ref{screenedflux} can be used to  estimate both the jump in $\Phi_{self}$ and the inductance $L$ of the ring. In the critical state at $B_a = B_p$ the ring current density is $J_c$ everywhere, so the total current is $J_c d w$. Since the flux due to the applied field is exactly cancelled by the supercurrents, we have $\Phi_{self} = B_p A_{r < r_p}$, and we find the inductance $L = B_p A_{r < r_p} / J_c d w = 4$~nH. From the first jumps we estimate  $\Delta \Phi_{self} / \Phi_{self} = 0.2$. Notice that we now are concerned with the self-flux of the ring and consequently the contribution from the applied field must be subtracted in the curves in figure \ref{screenedflux}. The width $w$ of the resistive region can be estimated  from the core width of the dendrites. Barkov \emph{et al} \cite{Barkov2003} found that typically  $w = 10^{-5}$~m.  With the normal state resistivity  used in the temperature estimate we obtain  $\delta t_2 \approx 4 \cdot 10^{-7}$~s.  We stress that this estimate represents a lower bound on the true timescale. This is because  the temperature is below $T_c$  during part of  the process, with a flux flow resistivity $\rho_f$ which may be orders of magnitude smaller than the $\rho_n$ we used to estimate the duration.  However, most of the current change should occur when the resistivity is highest, i.e. $\rho = \rho_n$, and our $\delta t_2$ based on a constant resistivity is expected to be  accurate. 

 In ignoring the term in equation (\ref{eq:heatequation}) which describes heat diffusion into the surrounding superconducting material, we neglect a possible source of heat loss which would lead to a lower maximum temperature. With such a term equations (\ref{LR}) and (\ref{eq:heatequation}) can be solved numerically to yield the complete temperature evolution as well as the width of the resistive region with only a single input from our experiments, namely the change in flux $\Phi_{r<r_p}$. However, the calculation requires  temperature dependent model parameters and detailed initial conditions for phase 2, which are known only approximately. We therefore believe our approximate solution is as reliable as such a full scale model would be.

\section{Current profiles due to perforations}
\label{sec:simulations}

In the discussion up to this point we have disregarded the non-uniform change in the current distribution which was pointed out in  section \ref{sec:results}. In  this section we  calculate the current distribution  resulting from a perforation within the resistive channel model, and find that key features of the observations can be reproduced. The calculation scheme    consists in computing the critical state of the ring up to some pre-defined applied field where a counter current is injected,  reducing the total current and changing the current distribution. The trick is similar to the one used in recent work on flux focusing in SQUIDs with a narrow slit \cite{brojeny:174514,brandt:024529}. It is motivated by the desire to retain the tractable cylindrical symmetry: the counter current provides  a simple way to model the \emph{effect} of the resistive slit without sacrificing the symmetry, and without considering the dynamical details of the process. 

The counter current distribution   should leave the field inside the superconducting material unchanged in line with the observation from figure \ref{perforationimages} (b), a constraint which   is fulfilled by a  transport Meissner current for a ring. Of course, it is difficult in practice to  impose a pure transport current on a ring:  therefore  this problem has received no attention in the literature so far, in contrast to the analogous problem of a  flat strip with a transport current  \cite{0022-3727-3-4-308,Zeldov1994:1,Brandt1993}. 

The calculations of the field and current distributions are inspired by the works of Brandt \cite{PhysRevB.50.4034,PhysRevB.54.4246,Brandt1997,PhysRevB.58.6506}, which details an efficient numerical method to model the electrodynamics in flat superconductors. The starting point is  Maxwells equations, written as $\nabla ^2 \mathbf{A} = \mu_0 \mathbf{J}$, and a suitable material law  describing the superconducting material, $E = E_c (J/J_c)^n$; $E$ is the electric field, $J$ is the current, $J_c$ is the critical current of the superconductor, $E_c$ is a critical electric field, and $n$ is the creep exponent. While this $E-J$ law describes flux creep, it is conveniently used in numerical calculations to describe the critical state if the exponent $n$ is high enough. These two equations lead to an equation of motion for the current (see refs \onlinecite{PhysRevB.54.4246,Brandt1997} for details) which can be written in cylindrical coordinates as

\begin{equation}
  \dot{J}(r) = \int_{r_{0}}^{r_{1}} Q^{-1}(r,r') \left( J(r')^n \frac{E_c}{J_c d} - \frac{r}{2} \dot{B}_a \right) dr'
  \label{Jeqnmotion}
\end{equation}

$B_a$ is the applied field. The integral kernel $Q(r,r')$ is given by

\begin{equation}
  Q(r,r') = \frac{1}{2 \pi} \int_{0}^{\pi} \frac{r'cos \phi } {\sqrt{r^2 + r'^2 + 2rr'cos \phi}} d\phi
   \label{kernel}
\end{equation}

and relates the magnetic vector potential $A(r)$ everywhere to the current $J(r')$ in the ring via

\begin{equation}
  A = \int_{r_{0}}^{r_{1}} Q(r, r') J(r') dr'
  \label{A_J}
\end{equation}

We integrate equation (\ref{Jeqnmotion}) numerically in a straightforward manner.  The kernel is found by computing equation (\ref{kernel}) at discrete points using a non-equidistant grid with high grid density near the sample edges \cite{Brandt1997}.  The ensuing matrix is then inverted to obtain $Q^{-1}(r,r')$ required in equation (\ref{Jeqnmotion}). A technical challenge  with this formulation is that  $Q(r, r')$ diverges for $r = r'$, but this can be handled by carefully selecting finite diagonal elements  (see the appendix for more details).

The magnetic  field distribution is  calculated from $J(r)$ by first
computing the vector potential from equation \ref{A_J}, and inserting
this into $B(r) = \partial A(r) / \partial r + A(r) / r + B_a$.  The computed flux $\Phi_{r < r_p}  = \int_S B dS = 2\pi \int_0^r B(r') r' dr'$ from  a simulation is shown in figure \ref{screenedflux}, and agrees well with the experimental curves. To achieve this close matching it is necessary to include a field dependent critical current in the simulations. We used  a Kim model $J_c = J_{c0} / (1 + B / B_K) $, with  $B_K = 80$~mT. This value is only a little higher than expected  based on previously reported \cite{Kim2001} magnetisation data on the same sort of samples.   

In order to complete the perforation model we also need the distribution of a Meissner transport current in this cylindrical symmetry. This may be found by inverting  equation (\ref{A_J}) to find the current $J(r)$  from a magnetic potential $A(r)$. With  $A(r) = K/r$ the magnetic field inside the superconducting material is zero, so the desired current distribution can be expressed as 

\begin{equation}
  J_{tr}(r) = K \int_{r_{0}}^{r_{1}} \frac{Q^{-1}(r, r')}{r'} dr'
  \label{Jtr}
\end{equation}

It is convenient to choose the constant $K$ such that the total transport current equals the total shielding current at perforation, i.e. that $\int J_{tr} (r) dr = \int J_{before} (r) dr$. The net current just after a perforation can then be written 

\begin{equation}
  J_{after}(r) = J_{before}(r) -  \alpha J_{tr}(r)
  \label{Jtot}
\end{equation}

\begin{figure}
  \includegraphics[width=\halfwidth]{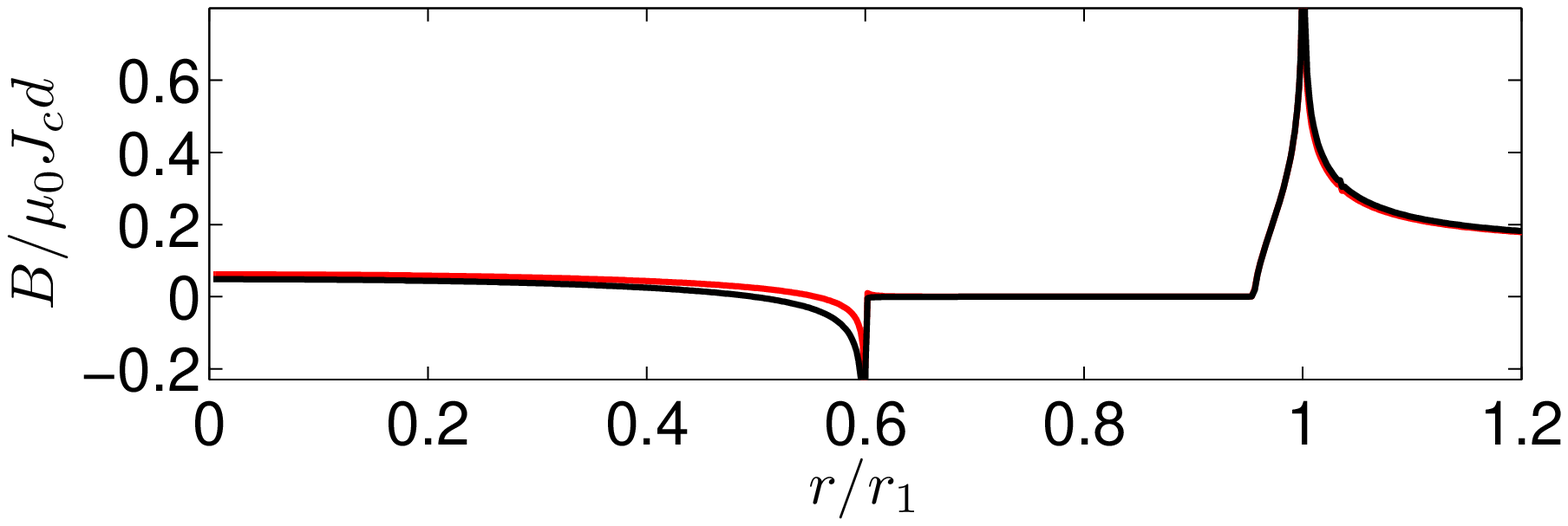} \\
  \includegraphics[width=\halfwidth]{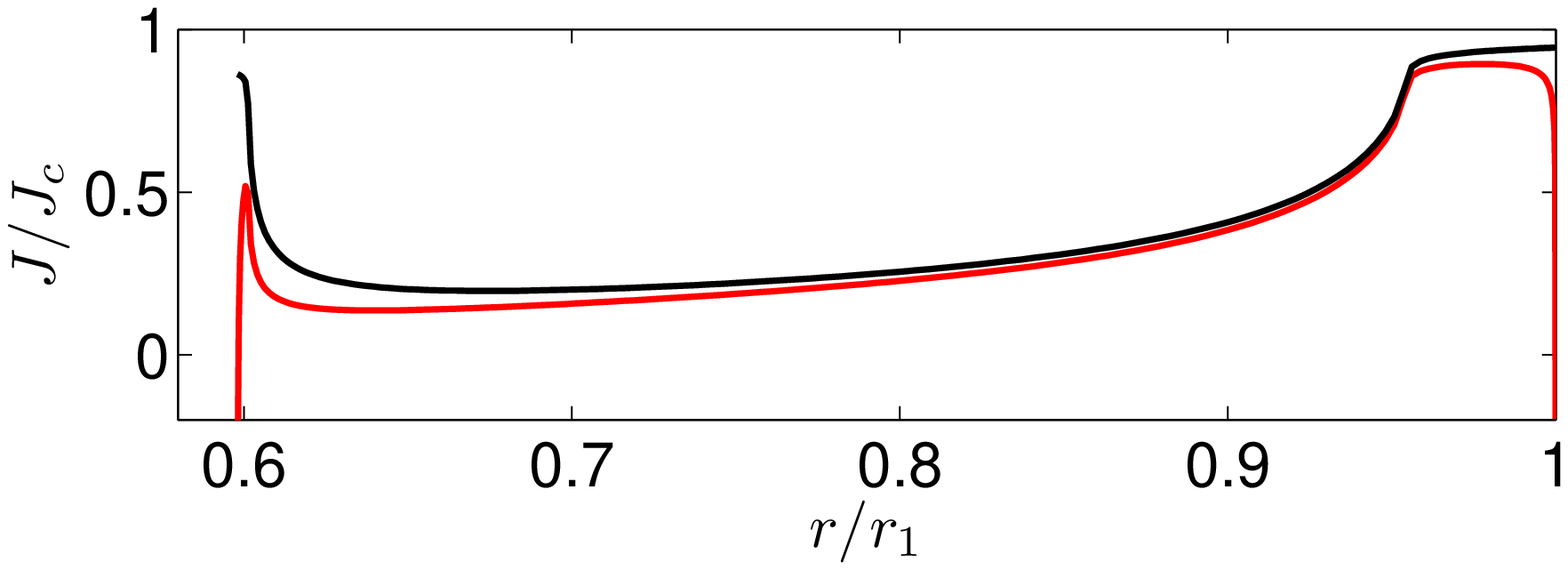}
  \caption{Field (top) and current profiles   computed just prior to and just after a simulated perforation. The perforation field was 0.15 $\mu_0 J_c d$, the creep exponent $n=30$, and the blocking parameter $\alpha=0.13$, corresponding to typical values from our experiments (see figure \ref{block_vs_perf}). }
  \label{fig:simprofiles}
\end{figure}

The parameter $\alpha \in [0,1]$ determines the reduction of the total ring current as a result of a perforation. This $\alpha$ is analogous to the relative change in flux $\alpha ' = \Delta \Phi_{self} / \Phi_{r < r_p}$ in the simplified model described in section \ref{sec:peaktemperature}.  We artificially inject this current at
a predetermined applied field, and subsequently let the simulation
proceed normally, i.e. according to equation \ref{Jeqnmotion} again. Figure \ref{fig:simprofiles} shows field and current profiles just before and shortly  after a perforation. We observe that most of the current change occurs near the inner edge. In the Meissner region there is some noticeable reduction in the current closer to the inner flux front, but towards the outer flux front there is hardly any change at all. Near the outer edge, though, the current has decreased considerably. These features compare favourably with the experimental current profiles shown in  figures \ref{perforationimages} and \ref{fig:currentevolution}.  

\begin{figure}
  \includegraphics[width=\halfwidth]{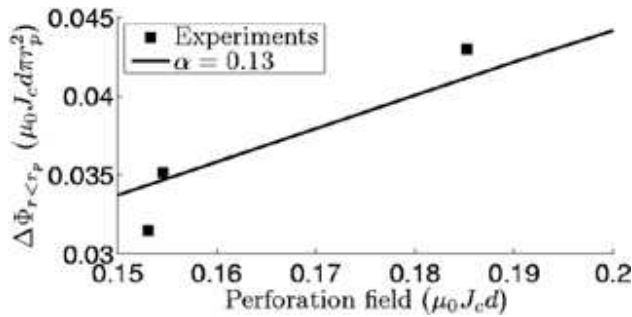}
  \caption{Flux jump size versus perforation field from simulations
    and experiments. The filled squares show experimental first jumps, and the solid line shows results from simulations with $\alpha = 0.13$. The simulations reproduce the experimental fact that the current is reduced by a smaller amount than $\Phi_{r < r_p}$}
  \label{block_vs_perf}
\end{figure}

In figure \ref{block_vs_perf}
we show computed values of the flux jumps size versus the perforation
field for different values of the blocking parameter $\alpha$. Also included in
the plot is the first jump in some experiments. The plot shows that the total ring current is reduced by approximately 13 \% by perforations. This is considerably smaller than the reduction we would infer from the plots of $\Phi_{r < r_p}$ in figure \ref{screenedflux} using $\Phi = LI$, where the relative change in the flux is $\approx 0.2$ for the first jumps. This is of course due to the fact that most of the current change occurs near the inner edge where the effect on the flux in the central hole is greatest. The value $\alpha = 0.13$ also is more in line with the plot of total current in figure \ref{fig:currentevolution}, which shows a relative change in current of $\approx 0.16$ for the first jump. 

\section{Conclusion}

We have proposed a method to estimate the temperature rise during dendritic avalanches in superconducting thin films. This is accomplished using MO imaging to measure the flux change in the central hole of a superconducting MgB$_2$ ring caused by avalanches stretching from the outer to the inner edges of the ring. Our model assumes that the dendrite tip creates a short-lived, resistive  heated channel as it crosses the ring, injecting flux into the central hole. From this model we are able to estimate the peak temperature during flux injection to be $T_{max} = 2.5T_c = 105$~K, the duration of process to be $\delta t = 4 \cdot 10 ^ {-7}$~s, and calculate the current profiles before and after injection events. 

\begin{acknowledgments}
We thank Paul Leiderer for doing ultrafast MO imaging on MgB$_2$ samples following our request.
{\AA}AFO, DS and THJ greatfully acknowledge the financial support from  FUNMAT@UiO and the Research Council of Norway.
\end{acknowledgments}

\appendix

\section{Stability of the numerical calculations}
\label{sec:stability}

The stability of the numerical calculations described in section \ref{sec:simulations} hinges on a careful selection of finite replacement elements for the diverging diagonal terms in the kernel, equation (\ref{kernel}).  The problem  has been dealt with in various ways previously \cite{brandt:184509,PhysRevB.58.6506,PhysRevB.54.3530,PhysRevB.50.4034}. One way to obtain diagonal elements that work is the following: With an appropriate grid $r_i$ and corresponding weights $w_i$ one can integrate $Q$ numerically over the ring width: 

\begin{equation}
  \label{eq:numint}
  \int_{r_0}^{r_1} Q(r, r') dr' \approx \sum_{i} Q_{ij} w_i
\end{equation}

The right hand side  can be split into diagonal and off-diagonal terms, and the equation can be rewritten as

\begin{equation}
  \label{eq:diagonal}
  Q_{ii} w_i = \int_{r_0}^{r_1} Q(r, r') dr'  - \sum_{i \neq j} Q_{ij} w_i
\end{equation}

The integral of $Q(r, r')$ over $r'$ can be carried out by changing the order of integration of $\phi$ and $r'$. The $r'$-integration can be done analytically; the result is a smooth integrand for the $\phi$ integration except for a singularity at $\phi = \pi$, but the integral is well defined over the entire interval $\phi \in [0, \pi]$. Equation (\ref{kernel}) can be used to compute the off-diagonal elements on the right hand side, and the resulting set of diagonal elements  give  stable and accurate computations.

An additional numerical challenge comes from the fact that  $J_{tr}$ obtained from equation (\ref{Jtr}) diverges at the edges. One way to cope with this for a regular Bean model (no field dependent $J_c$) is to modify the $E-J$ law so that for current above $|J_c|$ there is an ohmic  resistivity $\rho_f$, in mathematical form  that

\begin{equation} \nonumber E = \left\{
    \begin{array}{ll}
      \rho_f J & \mbox{ if $J > J_c $} \\
      E_c \arrowvert \frac{J}{J_c} \arrowvert ^ n \mathrm{sgn} (J) & \mbox{ if $J \leq J_c $}
    \end{array}
  \right.
\end{equation}

With this small modification the supercritical currents quickly relax in the simulations. 

The resistivity does not necessarily have a physical meaning in this context since its main purpose is to stabilise the simulation when the current reaches extremely large values. Still, such an ohmic resistivity is realistic in superconductors for currents larger than the critical current, and the use of the modified $E-J$ law is reasonable also from a physical point of view. We use a $\rho_f/\mu_0 = 0.001$~s$^{-1}$. The value will in reality
depend on magnetic field since it captures the motion of vortices -
the more vortices moving will generate a larger electric field acting
against the current. However, for the computations the exact value is
not crucial provided it is small.

\end{document}